\documentclass[twocolumn,showpacs,epsfig,preprintnumbers,amsmath,amssymb]{revtex4}
\usepackage{times,mathptmx}
\usepackage{amsmath,units,amssymb}
\usepackage{mathrsfs}
\usepackage{units,xspace}
\usepackage{graphicx}
\usepackage{dcolumn}
\usepackage{bm}
\def \be{\begin{equation}}
\def \ee{\end{equation}}
\def \bea{\begin{eqnarray}}
\def \eea{\end{eqnarray}}

\def \ben{\begin{enumerate}}
\def \een{\end{enumerate}}

\newcommand{\kT}{k_{\rm B}T}
\newcommand{\MSDa}{{\rm MSD}^{\rm 1P}}
\newcommand{\MSDb}{{\rm MSD}^{\rm 2P}}
\newcommand{\la}{{\langle \hspace{0.1em} l \hspace{0.1em} \rangle}} 
\newcommand{\cF}{\mathscr{F}}

\begin{document}

\title{Extracting the dynamic correlation length of actin networks from microrheology experiments}
\author{Adar Sonn-Segev}

\affiliation{Raymond \& Beverly Sackler School of Chemistry, Tel Aviv
  University, Tel Aviv 6997801, Israel}

\author{Anne Bernheim-Groswasser}

\affiliation{Department of Chemical Engineering, Ilse Kats Institute
  for Nanoscale Science and Technology, Ben Gurion University of the
  Negev, Beer-Sheva 84105, Israel}

\author{Yael Roichman}
\email{roichman@tau.ac.il}

\affiliation{Raymond \& Beverly Sackler School of Chemistry, Tel Aviv
  University, Tel Aviv 6997801, Israel}
\date{\today}

\begin{abstract}

The mechanical properties of polymer gels based on cytoskeleton proteins (e.g. actin) have been studied extensively due to their significant role in biological cell motility and in maintaining the cell's structural integrity. Microrheology is the natural method of choice for such studies due to its economy in sample volume, its wide frequency range, and its spatial sensitivity. In microrheology, the thermal motion of tracer particles embedded in a complex fluid is used to extract the fluid's viscoelastic properties. Comparing the motion of a single particle to the correlated motion of particle pairs, it is possible to extract viscoelastic properties at different length scales.  In a recent study, a crossover between intermediate and bulk response of complex fluids was discovered in microrheology measurements of reconstituted actin networks. This crossover length was related to structural and mechanical properties of the networks, such as their mesh size and dynamic correlation length. Here we capitalize on this result giving a detailed description of our analysis scheme, and demonstrating how this relation can be used to extract the dynamic correlation length of a polymer network. We further study the relation between the dynamic correlation length and the structure of the network, by introducing a new length scale, the average filament length, without altering the network's  mesh size. Contrary to the prevailing assumption, that the dynamic correlation length is equivalent to the mesh size of the network, we find that the dynamic correlation length increases once the filament length is reduced below the crossover distance.
\end{abstract}

\maketitle
\section{Introduction}

Complex fluids are intriguing materials, both from the structural and the mechanical point of view. Comprised of at least two components, these fluids contain mesoscopic structural features on the scale of nanometers to millimeters \cite{WittenBook}. As a result their mechanical response to perturbations is both elastic-like and fluid-like in nature. Conventionally, complex fluids are characterized mechanically by bulk rheology\cite{LarsonBook}. Complex fluids of biological origin, which are not readily available in large quantities, are usually characterized using a more material economic technique, microrheology, which uses the motion of embedded tracer particles observed by optical microscopy to extract the material properties\cite{Mason1995,Mason1997,Gittes1997a,Gittes1997b,Crocker2000,Squires2010,Levine2000}. Another advantage of microrheology is its ability to characterize the viscoelastic properties of these fluids on different length scales\cite{Chen2003,Starrs2003,Liu2006}. Utilizing this trait of microrheology, we recently showed that the mechanical properties of an example complex fluid (actin networks) change from bulk to intermediate behavior below a characteristic crossover length ($r_c$)\cite{Sonn2014}. This new length scale depends both on structural features of the material as well as on its local and bulk viscoelastic properties. The crossover length, $r_c$, can be related to the dynamical correlation length, $\xi_d$, of the complex fluid. For polymer networks, $\xi_d$, which is the length scale over which dynamical correlations decay in the network, is considered to be related to the mesh size\cite{deGennes1976a,deGennes1976b}, and is commonly measured by dynamic light scattering, requiring large sample volumes. Measuring $\xi_d$ with microrheology offers a means to connect mechanical properties of a polymer networks to their structure using microscopic quantities. 

Polymer networks made of cytoskeleton proteins have been thoroughly studied in an effort to understand their biological role in the cell \cite{Kasza2007,Gardel2008,Mofrad2009,Stricker20109,Fletcher2009,Chen2010,MacKintosh2010,Smith2010,Wen2011}. The most researched of which is actin, which is the focus of this paper. We study the spatial dependence of the viscoelastic properties of thermally equilibrated F-actin networks, and their relation to the networks' structure. We start by outlining our generalized analysis scheme of microrheology experiments and its application to reconstituted actin networks of different mesh size. We then demonstrate how to extract the viscoelastic and structural properties of the networks, regardless of tracer particle size (i.e., its size relative to the mesh size). We proceed to explore the dynamical correlation length's relation to the networks' mesh size, and investigate how $\xi_d$ is affected by the introduction of another relevant length scale, the average filament length~$\la$. Finally, we examine the relation between the viscoelastic plateau modulus and the dynamic correlation length of the gels.

\section{Experimental}

We use entangled F-actin networks as a model viscoelastic fluid. The rheological properties of this system have been studied extensively both experimentally and theoretically \cite{Palmer1999,Crocker2000,Gardel2003,Shin2004,Liu2006,Atakhorrami2014,Levine2000}. F-actin gels are well described as networks of semiflexible polymers, and their mesh size, $\xi_s=0.3/\sqrt{c_A}$, is easily controlled through monomer concentration  $c_A$ ($c_A$ in mg/ml and $\xi_s$ in $\mu$m \cite{SCHMIDT1989}).

G-actin is purified from rabbit skeletal muscle acetone powder \cite{Spudich1971}, with a gel filtration step, stored on ice in G-buffer (5 mM Tris HCl, 0.1 mM CaCl$_2$, 0.2 mM ATP, 1 mM DTT, 0.01\% NaN$_3$, $p$H 7.8) and used within two weeks. The concentration of the G-actin is determined by absorbance measured using UV/Visible spectrophotometer (Ultraspec 2100 pro, Pharmacia) in a cuvette with a 1 cm path length and extinction coefficient $\epsilon_{290}=26,460$ M$^{-1}$cm$^{-1}$.
Polystyrene colloids with diameters of $a=0.245, ~0.55~\mu$m (Invitrogen Lots \#1173396 and \#742530 respectively) are pre-incubated with a 10 mg/ml BSA solution to prevent non specific binding of protein to the bead surface \cite{Valentine2004}. The average filament length, $ \la $, is controlled by addition of capping protein (CP). Actin polymerization is initiated by adding G-actin in various concentrations, CP and beads to F-buffer solution (5 mM Tris HCl, 2 mM MgCl$_2$, 0.05 M KCl, 200 $\mu$M EGTA, 1 mM ATP) and mixing gently for 10 sec. Mesh size is varied by changing G-actin concentration between $c_A~=~0.46-2$~mg/ml, corresponding to  $\xi_s=0.44-0.21$~$\mu$m (at fixed $ \la=13~\mu$m). The average filament length is varied, at constant actin concentration ($\xi_s=0.3~\mu$m), by changing the concentration ratio of actin/CP. Filament length distribution is roughly exponential \cite{Xu1999}. We estimate $ \la~=~2~-~13~\mu$m assuming CP determines the number of actin nucleation sites \cite{Janmey1986,Xu1999,Liu2006}.

Immediately after polymerization the samples were loaded into a glass cell,  $150~\mu$m high, and sealed with grease. The glass surfaces were coated with methoxy-terminated PEG to prevent binding of the network to the glass.  After equilibrating for 30 min at room temperature, samples were imaged at a plane distanced from the cell walls with an epi-fluorescence microscope (Olympus IX71), at $\lambda=605$ nm, with 60x oil, and 40x air objectives for $a=0.245~\mu$m and $a=0.55 ~\mu$m, respectively. We recorded the motion of approximately 100 particles in the field of view using a CMOS video camera (Gazelle, Point Gray) at a frame rate of 70 Hz with an exposure time of 0.003 sec. To insure high signal to noise ratio of two-particle displacement correlation measurements, we used data from approximately $8\cdot 10^5$ frames per experiment. Particle tracking was done using conventional algorithms with accuracy of at least 13 nm \cite{Crocker1996}.

\section{Generalized microrheology and the dynamic correlation length}

\subsection{Microrheology at intermediate length scales}
\begin{figure}[h!]
\includegraphics[scale=0.30]{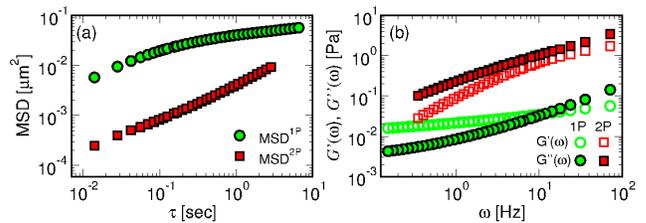}
\caption{Microrheology of entangled F-actin networks. (a) $\MSDa$
  (green) and $\MSDb$ (red) as a function of lag time, for $\xi_s=0.3~\mu$m,  $a=0.245~\mu$m and $\la =13~\mu$m. (b) The storage modulus, $G'(\omega)$
  (open symbols), and loss modulus, $G''(\omega)$ (filled symbols),
  extracted from the $\MSDa$ (green) and $\MSDb$ (red) curves of panel
  (a).}
\label{fig:msds}
\end{figure}

Conventional microrheology is concerned with characterizing the mechanical properties of a complex fluid by analyzing the diffusion of tracer particles embedded in it \cite{Mason1995,Mason1997,Gittes1997a,Gittes1997b,Crocker2000,Squires2010,Levine2000}. We concentrate on the passive variants of the technique \cite{Squires2010} relating the thermal fluctuations of the tracer particles to the viscoelastic properties of the characterized fluid, using both one point (1P) and two point (2P) microrheology. In 1P microrheology the generalized Stokes Einstein relation (GSER) is used to connect the ensemble averaged mean-squared displacement of tracer particles,  MSD$^\text{1P}\equiv \langle \Delta x^2(\tau)\rangle$ (Fig.~\ref{fig:msds}(a)) to the viscoelastic moduli, G'($\omega$) and G''($\omega$) (Fig.~\ref{fig:msds}(b)) \cite{Mason1995,Crocker2007,Squires2010}. 

This technique probes only the local environment of the tracer particle, which is the microscopic volume explored by the particle within the experimental time scale. Consequently, it is well established that 1P microrheology of actin networks underestimates the bulk viscoelastic moduli\cite{Crocker2000,Gardel2003,Shin2004,Liu2006,Atakhorrami2014}. 2P microrheology was developed to address this issue, by looking at the average correlated diffusion of two distanced particles. Specifically, one measures the ensemble-averaged longitudinal and transverse displacement correlations of particle pairs as a function of inter-particle distance $r$ and lag time $\tau$ \cite{Crocker2000}:
\begin{eqnarray}
  D_\parallel(r,\tau) &=& \langle\Delta r^i_\parallel(t,\tau) \Delta r^j_\parallel(t,\tau)
  \delta(r-R^{ij}(t))\rangle \nonumber\\
  D_\perp(r,\tau) &=& \langle\Delta r^i_\perp(t,\tau) \Delta r^j_\perp(t,\tau)
  \delta(r-R^{ij}(t))\rangle,
\label{drr}
\end{eqnarray}
where $\Delta r^i_\parallel(t,\tau)$ ($\Delta r^i_\perp(t,\tau)$) is
the displacement of particle $i$ during the time between $t$ and
$t+\tau$, projected parallel (perpendicular) to the line connecting
the pair, and $R^{ij}(t)$ is the pair separation at time $t$. At
sufficiently large distances both correlations decay as $r^{-1}$,
$D_\parallel\simeq A(\tau)/r$ and $D_\perp\simeq A(\tau)/(2r)$. The
common practice is to use this asymptote to define a `two-point
mean-squared displacement', $\MSDb(\tau) \equiv 2A(\tau)/(3a)$~\footnote[4]{We use the one-dimensional forms of the MSD's.},
and extract from it the viscoelastic moduli using again the GSER
\cite{Crocker2000}.

Figures~\ref{fig:msds} (a) and (b) show the 1P and
2P MSD's measured in an actin network ($\xi_s=0.3~\mu$m), and the moduli
extracted from them.  The viscoelastic properties obtained from the
two approaches are significantly different, demonstrating the much
softer local environment probed by the 1P technique, as compared to
the bulk response probed by the 2P one.  These results are in accord,
both qualitatively and quantitatively, with previous studies on
F-actin networks \cite{Crocker2000,Gardel2003,Liu2006}.

\begin{figure}[h!]
\includegraphics[scale=0.30]{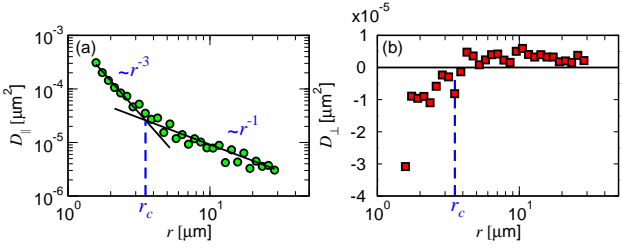}
\caption{(a) Longitudinal and (b) transverse displacement correlations as a function of particle separation, $r$, at lag time $\tau=0.014~\unit{sec}$ for $\xi_s=0.3~\mu$m,  $a=0.245~\mu$m and $\la =13~\mu$m. The crossover distance $r_c$ (blue dashed line) is defined at the intersection of the fitted dominant ($r^{-1}$) and subdominant ($r^{-3}$) power-law decays of $D_\parallel$.}
\label{fig:Drq}
\end{figure}

We have recently shown \cite{Sonn2014} that the inter-particle distance at which the bulk response sets in is much larger than would intuitively be expected. For example, in our experiments (Fig.~\ref{fig:Drq} (a) and (b)) a crossover to an intermediate regime is observed at $r_c= 3.5~\mu$m, which is an order of magnitude larger than the network mesh size, $\xi_s=0.3~\mu$m, and the tracer particle's radius, $a=0.245~\mu$m. The detailed theoretical description of the viscoelastic behavior of complex fluids at intermediate length scales, below $r_c$, is given elsewhere \cite{Sonn2014,Diamant2014}. Simply stated, a particle moving within a fluid disturbs it in two ways: it generates a momentum perturbation that spreads in the bulk, and displaces mass locally \cite{Diamant2007}. These two contributions can be expanded in terms of inter-particle distance and depend on the bulk and local viscosity respectively. For complex fluids in which the local environment is much softer than the bulk, the leading terms in the mobility expansion are \cite{Sonn2014,Diamant2014}: 

\begin{eqnarray}
M_{ \parallel } (r, \omega ) &=& \frac{1}{4\pi\eta_br} + \frac{a^2g(\xi_d/a)}{2\pi\eta_\ell r^3} \label{Eq:mobr} \\
M_{\perp}(r, \omega) &=& \frac{1}{8\pi\eta_br} - \frac{a^2g(\xi_d/a)}{4\pi\eta_\ell r^3},
\label{Eq:mob} 
\end{eqnarray}
where $\eta_b$ ($\eta_\ell$) corresponds to the bulk (local) viscosity and the function $g(\xi_d/a)$ arises from calculating the fluid response to a forced rigid sphere of finite radius $a$.  The first term,  the dominant response, arises from momentum conservation, while the second term, the sub-dominant response, describes mass transfer. At intermediate distances ($r\lesssim r_c$) the viscoelastic properties of a complex fluid are governed by the subdominant term \cite{Sonn2014,Diamant2014}. Equations (\ref{Eq:mobr}) and (\ref{Eq:mob}) imply that the intermediate response should decay as $1/r^3$ in the longitudinal direction, and exhibit negative correlation in the transverse one.
As a result the crossover between the asymptotic, dominant response in the longitudinal direction to the intermediate, subdominant one should appear at a distance: 

\begin{equation}
  r_c = a [2(\eta_b/\eta_\ell)g(\xi_d/a)]^{1/2},  
\label{rc}
\end{equation}
where $g(x)$ is a material specific function that satisfies the asymptotic conditions\cite{Sonn2014,Diamant2014}: $g(x\rightarrow\infty)=x^2$, and $g(x~\rightarrow~0)=~1$.
 
The displacement correlation, $D_\parallel$, can be related to the mobility $M_\parallel$, using the fluctuation-dissipation theorem:
\begin{equation}
D_\parallel(r, \omega)=-(2k_\text{B}T/\omega^2)M_\parallel(r, \omega),
\end{equation} 
where $k_\text{B}T$ is the thermal energy. To minimize data manipulation the analysis is applied on the time (rather than frequency) domain and thus the expected expression for $D_\parallel(r, \tau)$ is:
\begin{equation}
D_\parallel(r, \tau)=\frac{A(\tau)}{ r}+ \frac{B(\tau)}{r^3}
\label{Eq:DP}
\end{equation}
where
\begin{eqnarray}
A(\tau)&=&\frac{k_\text{B}T}{2\pi}\cF^{-1} \left\lbrace \frac{1}{-\omega^2\eta_b}\right\rbrace \label{Eq:A}\\
B(\tau)&=&\frac{k_\text{B}T }{\pi}a^2g\left(\xi_d/a\right)\cF^{-1}\left\lbrace \frac{1}{-\omega^2\eta_\ell}\right\rbrace
\label{Eq:B}
\end{eqnarray}

where $\cF^{-1}$ denotes the inverse Fourier  transform. The crossover distance in the time domain is then given by:
\begin{equation}
r_c(\tau)=[B(\tau)/A(\tau)]^{1/2}.
\label{Eq:R_tau}
\end{equation}

\subsection{Dynamic correlation length measurement}

One outcome of the preceding theory is that the dynamic correlation length of a complex fluid can be extracted from microrheology experiments, provided that: (1) the functional form of $g(x)$ is known, (2) the crossover distance (Eq.~(\ref{Eq:R_tau})) is experimentally observed, and (3) the bulk and local viscosity are measured. 
We start our analysis by expressing $A(\tau)$ and $B(\tau)$ in terms of $\MSDb$ and $\MSDa$ respectively,
\begin{eqnarray}
A(\tau)&=&3a\MSDb/2\nonumber\\
B(\tau)&=& 3a^3g(\xi_d/a)\MSDa.
\end{eqnarray}

To this end we assume that the local viscosity is a function of time and is related to the $\MSDa$ by the fluctuation-dissipation theorem $\MSDa(\tau)
= [\kT/(3\pi a)]\cF^{-1}\{(-\omega^2\eta_\ell)^{-1}\}$.
The bulk viscosity is related to the $\MSDb$ in a similar manner , and is given by $A(\tau)$ (Eq.~(\ref{Eq:A})). 
\vspace{0.5em}

\begin{figure}[h!]
\includegraphics[scale=0.30]{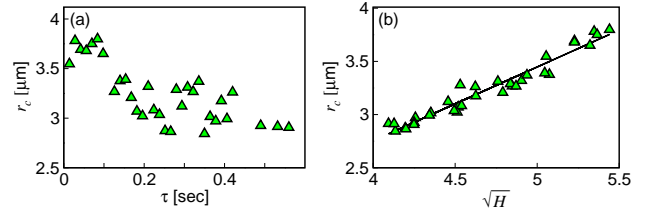}
\caption{Crossover distance as a function of (a) lag time, and (b)
  square root of $H(\tau)$, the experimental function characterizing
  the bulk to local viscosity ratio. Parameter values are the same as in Fig.~\ref{fig:msds}; $a=0.245~\mu$m, $\xi_s=0.3~\mu$m and $\la = 13~\mu$m.}
\label{fig:figure2}
\end{figure}

Substituting these expressions into Eq.~(\ref{Eq:R_tau}) we have,
\begin{equation}
r_c=\left[  2 a^2g\left( \tfrac{\xi_d}{a}\right)\frac{\MSDa}{\MSDb} \right] ^{1/2}=\left[  2 a^2g\left(\tfrac{\xi_d}{a}\right)H(\tau) \right] ^{1/2}
\label{Eq:rcn}
\end{equation}
where we define:
\begin{equation}
H(\tau)\equiv\frac{\MSDa}{\MSDb}.
\label{Eq:H}
\end{equation} 
as the time dependent observable, and $g(\xi_d/a)$ the structural element to be characterized. 
The functional form of $g(x)$ for actin networks was derived using the two-fluid model of polymer gels\cite{deGennes1976a,deGennes1976b,Milner1993,Levine2000,Levine2001,Powers2008}, and reads\cite{Sonn2014,Diamant2014};
\begin{equation}
g(x) = x^2+x+1/3.
\label{Eq:g}
\end{equation}
\begin{figure}[h!]
\includegraphics[scale=0.32]{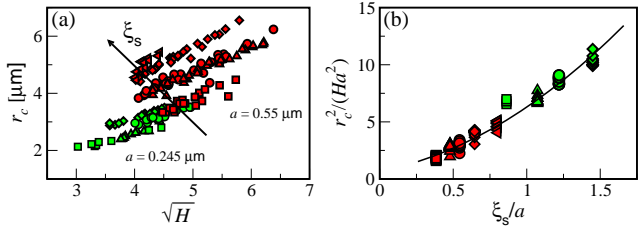}
\caption{Crossover distance for all experiments. (a) For all
  conditions $r_c$ is linear with $\sqrt{H}$ and increases with either
  $\xi_s$ or $a$. (b) All experimental results fall on a master curve
  once $r_c^2$ is normalized by $Ha^2$ and presented as a function of
  scale function given by Eq.~(\ref{Eq:g}) and $\xi_d=1.25\xi_s$. Red
  (Green) symbols correspond to $a=0.55$ ($0.245$) $\mu$m. Each symbol corresponds to a different mesh size: $\xi_s=0.21$ (squares), $0.26$ (triangles), $0.3$ (circles), $0.35$
  (diamonds), and $0.44$ $\mu$m (left triangles). The average filament length for all experiments was $\la=13~\mu$m. }
\label{fig:figure3}
\end{figure}

Recasting $r_c(\tau)$ as a function of $\sqrt{H(\tau)}$ reveals their linear dependence (Fig.~\ref{fig:figure2}(b)), as predicted theoretically in Eq.~(\ref{Eq:rcn}). This linear dependence holds for all our experiments, independent of tracer particle size and network mesh size (see Fig.~\ref{fig:figure3}(a)). Rescaling $r_c^2$ by $Ha^2$ and presenting it as a function of $\xi_s/a$ results in a collapse of our data on a single master curve shaped according to Eq.~(\ref{Eq:g}) (see Fig.~\ref{fig:figure3}(b))\cite{Sonn2014}. The only fitting parameter used to fit our data to Eq.~(\ref{Eq:g}), was the ratio $b\equiv\xi_d/\xi_s=1.25$. 
This result provides an experimental verification of the scaling function $g(\xi_d/a)$ derived using the two-fluid model for actin networks. Therefore providing a means to extract the dynamic correlation length from microrheology experiments.    

\begin{figure}[h!]
\includegraphics[scale=0.32]{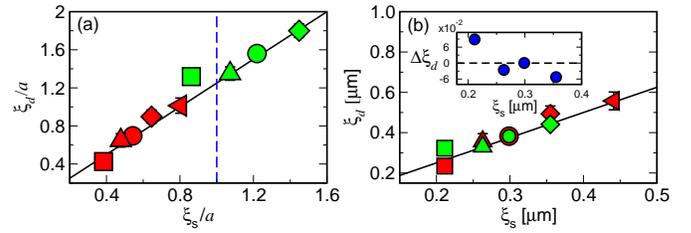}
\caption{Dynamic correlation length, $\xi_d$, extracted from $r_c$ and $H$ for networks with different mesh and particle sizes (see text for details). (a) $\xi_d$ scaled by $a$, particle radius, as a function of the scaled mesh size $\xi_s/a$. (b) $\xi_d$ as a function of $\xi_s$. Black line, in both figures, corresponds to $\xi_d=b\xi_s$, where $b=1.25$ is our fitting parameter.  Both $\xi_s/a>1$ and $\xi_s/a<1$ fall on the same line. Inset: Difference between $\xi_d$s extracted from the two particle sizes ($\Delta \xi_d$) as a function of $\xi_s$. Red (green) symbols correspond to $a=0.55$ ($0.245$) $\mu$m. Each symbol corresponds to a different mesh size: $\xi_s=0.21$ (squares), $0.26$ (triangles), $0.3$ (circles), $0.35$ (diamonds), and $\xi_s=0.44$ $\mu$m (left triangles).    }
\label{Fig:fig4}
\end{figure}

In Fig.~\ref{Fig:fig4} the measured dynamic correlation length is plotted versus the networks' mesh size for two different sizes of tracer particles. Both $\xi_d$ and $\xi_s$ are material properties and should not depend on the tracer particle size. The difference in the dynamic correlation length measured with the two different particle sizes is used to gauge its experimental error (see Fig.~\ref{Fig:fig4}(b)), which is of the order of $\langle\Delta\xi_d\rangle\simeq50$ nm. The fact that the relation between $\xi_d$ and $\xi_s$ is linear suggests that $\xi_d$ scales as the square root of actin concentration, as expected for semi-dilute polymer solution.  

\vspace{0.8em}

\section{Effect of filament length}
 
\begin{figure}[h!]
\includegraphics[scale=0.32]{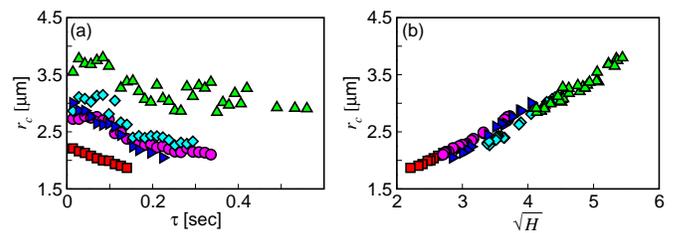}
\caption{Crossover distance, $r_c$, as a function of (a) lag time, and (b) square root of $H(\tau)$ for networks with different average filament length: $\la=2$ (red squares), $5$ (magenta circles), $8$ (blue right triangles), $10$ (cyan diamonds), and  $\la=13~\mu$m (green triangles). All networks were polymerized at the same concentration ($c_A=1$ mg/ml corresponding to $\xi_s=0.3~\mu$m). }
\label{fig:figure5}
\end{figure}

So far we have shown that the viscoelastic response of actin networks depends on the new emerging length scale $r_c$, rather than directly on the network mesh size or tracer particle size. In this section we introduce a new relevant length scale to the system, the average actin filament length, $\la$, which is controlled experimentally by introducing capping protein. We show that $\la$ affects the viscoelastic response of the networks if sufficiently decreased.
We study several networks made with the same actin monomer concentration but different average filament length, $\la=2,5,8,10,13~\mu$m. All of these systems create mechanically stable networks with a mesh size of $\xi_s=0.3~\mu$m, much smaller than the average filament length. While the mesh size is conserved in these systems it is not clear if their dynamic correlation length or their mechanical properties vary \cite{Liu2006}. 
Since the mesh size is the same in all of these gels and the average filament length is much larger than the mesh size,  we would naively expect the crossover length in these networks to be the same as well. In Fig.~\ref{fig:figure5} the crossover length scale of the different networks is examined. Surprisingly, even though the length scale depends linearly on the viscosities ratio (Fig.~\ref{fig:figure5}(b)) for each $\la$, it depends also on filament length (Fig.~\ref{fig:figure5}(a)). 

\vspace{0.5em}

\begin{figure}[h!]
\centering
\includegraphics[scale=0.4]{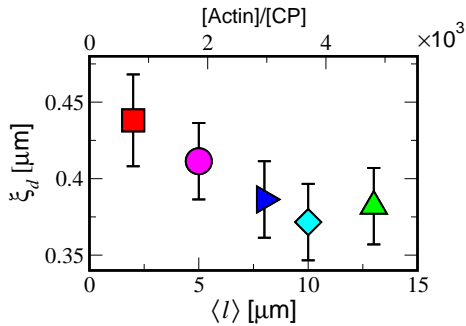}
\caption{Dynamic correlation length, $\xi_d$, as a function of the average filament length, $\la$ (bottom) and actin/CP concentration ratio (top). Actin concentration was held at 1 mg/ml, resulting in a $\xi_s=0.3~\mu$m, and $a=0.245~\mu$m. }
\label{fig6}
\end{figure}

A closer inspection of the data in Fig.~\ref{fig:figure5}(b) reveals that curves of different networks do not coincide, implying that the networks vary in dynamical properties. Since the functional form of $g(x)$ was calculated from the two-fluid model using a general, unspecified correlation length $\xi_d$, without any explicit reference to filament length, we can use it to extract $\xi_d$ of these networks. 
In Fig.~\ref{fig6} $\xi_d$ is plotted as a function of $\la$; for long filament length $\la>5~\mu$m $\xi_d$ does not depend on filament length, as expected. 
However, for shorter filaments, $\la=2,5~\mu$m, $\xi_d$ decreases with filament length. Note that the length scale below which $\xi_d$ is affected by filament length is of the order of $r_c$ and one order of magnitude larger than either $\xi_s$ and $a$. These results further the notion that $r_c$ is the length scale which is most relevant in determining explicitly the viscoelastic response of a complex fluid. The results also suggest that $\xi_d$ can be affected by other structural features of a polymer network, in addition to $\xi_s$, such as its dependence on $\la$ demonstrated here. 

\begin{figure}[h!]
\centering
\includegraphics[scale=0.32]{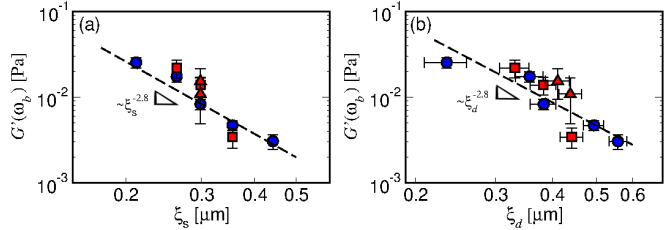}
\caption{The plateau elastic modulus, $G'(\omega_b) $, of all actin networks studied here, as a function of: (a) $\xi_s$ estimated from monomer concentration, and (b) $\xi_d$ extracted from measurements. Blue(red) symbols correspond to $a=0.55 (0.245)~\mu$m. Red squares correspond to $\la=13~\mu$m, and red triangles correspond to $\la=2,5~\mu$m.}
\label{Fig7}
\end{figure}

We characterize the viscoelastic properties of our networks in terms of the plateau modulus, $G'(\omega_b)$, with the lowest experimentally available frequency $\omega_b=0.14$~Hz, following Liu {\em et al.}\cite{Liu2006}. 
In Fig.~\ref{Fig7}(a) the plateau modulus, $G'(\omega_b)$, is plotted as a function of the actin network mesh size. As expected \cite{Isambert1996, Liu2006}, results from the various experiments fall on a single line showing a power law decay,  $G'(\omega_b)\propto \xi_s^\alpha$, with a power $\alpha\simeq-2.8$. However, the mesh size in these experiments is determined indirectly from the concentration of actin monomers used for gel preparation. We represent the results of Fig.~\ref{Fig7}(a) in terms of the directly measured correlation length $\xi_d$ (Fig.~\ref{Fig7}(b)). Here too all experiments fall on the same line with $\alpha\simeq-2.8$, even for networks with small filament length for which $\xi_d\neq b \xi_s$.  

\section{Conclusions}

In this paper we have presented a new method to extract the dynamic correlation length of complex fluids from microrheology measurements, and demonstrated it on a model system of entangled F-actin networks. This new technique is based on the observation of a crossover between the bulk and intermediate viscoelastic response of complex fluids in two point displacement correlations ($D_\|,D_\perp$). Using a generalized framework of analysis of microrheology, we show that the measured dynamic correlation length is related, but not identical, to the network mesh size. Specifically, when a third length scale is introduced into the problem, as demonstrated here with short filament lengths, $\xi_d$ depends on it as well as on $\xi_s$ (Fig.~\ref{fig6}). This latter result raises several questions: how is the dynamic correlation length related to the structure of a complex fluid, and consequently, how is it related to its viscoelastic properties. More detailed experiments are required to address these issues. The technique provided here presents a platform with which to characterize in more detail the dynamics of active complex fluids, such as biologically active actomyosin networks and chemically active self healing gels\cite{Cordier2008}.

\section*{Acknowledgments}

The authors are grateful to Haim Diamant for numerous illuminating discussions. This research was supported by the Marie Curie Reintegration Grant (PIRG04-GA-2008-239378), the Israel Science Foundation grant 1271/08, and by the US-Israel Binational Science Foundation grant 2008483. A. S.-S acknowledges funding from the Tel-Aviv University Center for Nanoscience and Nanotechnology. A. B.-G. acknowledges funding from the Israel Science Foundation (grant 1534/10).

\providecommand*{\mcitethebibliography}{\thebibliography}
\csname @ifundefined\endcsname{endmcitethebibliography}
{\let\endmcitethebibliography\endthebibliography}{}

\end{document}